\documentclass[%
prl%
 ,twocolumn%
 ,secnumarabic%
,tightenlines%
,amssymb, amsmath,nobibnotes, aps, prl, showpacs,
superscriptaddress]{revtex4}

\usepackage{docs}%
\usepackage{bm}%
\usepackage{cf2}%
\usepackage{graphicx}

\def\rv{{\bf r}}

\def\uv{{\bf u}}

\def\Fv{{\bf F}}

\def\Rv{{\bf R}}

\def\kB{k_{B}}
\def\zetav{\mbox{\boldmath $\zeta$}}
\def\gamv{\mbox{\boldmath $\gamma$}}

\def\etav{\mbox{\boldmath $\eta$}}

\def\la{\langle}
\def\ra{\rangle}

\def\bend{\mathsf{bend}}

\def\la{\langle}
\def\ra{\rangle}

\def\bend{\mathsf{bend}}

\def\bend{\mathsf{bend}}

\def\rv{{\bf r}}

\def\uv{{\bf u}}

\def\Fv{{\bf F}}

\def\Rv{{\bf R}}

\def\tauLp{\tau_{\textsf{diff},L_p}}

\expandafter\ifx\csname package@font\endcsname\relax\else
 \expandafter\expandafter
 \expandafter\usepackage
 \expandafter\expandafter
 \expandafter{\csname package@font\endcsname}%
\fi

\begin{document}

\title{Collapse of a semiflexible polymer in poor solvent}%

\author{Alberto Montesi}
\author{Matteo Pasquali}\email{mp@rice.edu}
\affiliation{Department of Chemical Engineering, Rice University,
6100 Main St., Houston, TX 77005, USA} \affiliation {The Kavli
Institute for Theoretical Physics, University of California, Santa
Barbara CA 93106, USA}
\author{F.C. MacKintosh}
\affiliation {The Kavli Institute for Theoretical Physics,
University of California, Santa Barbara CA 93106, USA}
\date{\today}%
\affiliation{Division of Physics and Astronomy, Vrije
Universiteit, 1081 HV Amsterdam, The Netherlands}

\begin{abstract}

We investigate the dynamics and the pathways of the collapse of a
single, semiflexible polymer in a poor solvent via 3-D Brownian
Dynamics simulations.  Earlier work indicates that the
condensation of semiflexible polymers generically proceeds via a
cascade through metastable racquet-shaped, long-lived
intermediates towards the stable torus state. We investigate the
rate of decay of uncollapsed states, analyze the preferential
pathways of condensation, and describe likelihood and lifespan of
the different metastable states. The simulation are performed with
a bead-stiff spring model with excluded volume interaction and
exponentially decaying attractive potential. The semiflexible
chain collapse is studied as functions of the three relevant
length scales of the phenomenon, i.e., the total chain length $L$,
the persistence length $L_p$ and the condensation length $L_0 =
\sqrt{k_B T L_p/u_0}$, where $u_0$ is a measure of the attractive
potential per unit length. Two dimensionless ratios, $L/L_p$ and
$L_0/L_p$, suffice to describe the decay rate of uncollapsed
states, which appears to scale as $(L/L_p)^{1/3} (L_0/L_p)$. The
condensation sequence is described in terms of the time series of
the well separated energy levels associated with each metastable
collapsed state. The collapsed states are described quantitatively
through the spatial correlation of tangent vectors along the
chain. We also compare the results obtained with a locally
inextensible bead-rod chain and with a phantom bead-spring model.
Finally, we show preliminary results on the effects of steady
shear flow on the kinetics of collapse.


\end{abstract}

\maketitle

\section{Introduction and motivation}

The conformation of individual polymer chains depends on the
solvent properties \cite{deGennesBook,BAH2,DoiEdwards}. In good
solvent the monomers prefer being surrounded by solvent molecules,
effectively repelling each other. This effect leads to a swollen
coil conformation for flexible polymers in good solvent. In poor
solvent, conversely, the monomers try to exclude the solvent and
effectively attract one another, and a flexible chain forms a
compact globule of roughly spherical shape to minimize the
contacts between monomers and solvent. The dynamics of the
coil-globule transition in flexible chains are well known
\cite{deGennes1985,Ostrovsky1994,Dawson1997,Halperin2000} and
involve the formation of a pearl necklace and the gradual
diffusion of large pearls from the chain ends
\cite{Chu1995,Abrams2002}.

Many polymers, however, exhibit substantial bending stiffness,
i.e., they are semiflexible. Such polymers can be described by the
worm-like chain model; examples include biopolymers (F-actin, DNA)
as well as synthetic polymers (xanthan, PBO, PPTA, kevlar).
Semiflexible polymers in good solvents form open, extended
structures, while in poor solvent the effective self-affinity of
the chain contrast this tendency and leads to chain collapse. The
collapsed state configurations and the pathways to their formation
are the result of the interplay between two opposing potentials:
the bending potential related to the chain stiffness and the
attractive potential due to the poor solvency of the environment.

A compact globule is energetically unfavorable for semiflexible
polymers because it involves highly bent states. The collapsed
ground state is instead a torus, which reduces the monomer-solvent
contacts without causing excessive bending penalty. Theoretical
analysis \cite{Grosberg1979,Bloomfield1997}, Monte Carlo
simulations \cite{Noguchi1996} and experimental evidences
\cite{Baeza1987,Martin2000} have detailed the stability, the
features and the packing of the collapsed tori. The dynamics of
the collapse of worm-like chains in poor solvent have been
investigated more recently
\cite{Schnurr2000,Noguchi2000,Sakaue2002}, strongly suggesting a
possible generic pathway for the collapse of semiflexible
polymers, featuring a series of long-lived, partially collapsed,
racquet-shaped intermediate states, before the eventual collapse
to a torus. Such intermediate states form an energetically driven
cascade of increasingly compact conformations with sharp
transitions between their energy levels.

In the present work, we verify this hypothesis and further explore
the effect of the relevant length scales on the phenomenon. The
natural length scale for the polymer stiffness is its persistence
length $L_p$, defined as the ratio of the bending stiffness
$\kappa$ to the thermal energy $k_B T$, which represents the
length along the chain for which tangent vectors remain correlated
in good solvent. The balance between attractive forces and bending
forces can be considered in term of the so-called ``condensation
length'' $L_0 = \sqrt{kT L_p/u_0}$, where $u_0$ is the value of
the attractive potential per unit length. We choose as time scale
for the collapse event the rotational diffusion time of a
perfectly rigid rod with $L = L_p$, $\tauLp = \zeta L^3_p / (72
k_B T)$, where $\zeta$ is the transverse friction coefficient
\cite{friction_coeff}. Therefore we use $\tauLp$ as unit of time
and $k_B T$ as unit of energy in presenting all the results of the
present work. We investigate the sequences and the likelihood of
different collapsed structures formed at various solvent quality
(described by $u_0$) and chain stiffness.

\section{Simulation details}\label{comp_mod}

The dynamics of the continuous wormlike chain are described by a
inertialess Langevin equation:
\begin{equation}
  \zetav \cdot \left \{
  \frac{\partial \rv}{\partial t} - \dot{\gamv}\cdot\rv \right \}
  =  -\frac{\delta U}{\delta \rv}
     + \etav \quad \label{Langevin}
\end{equation}
Equation \ref{Langevin} balances the hydrodynamic force exerted on
the chain by the surrounding fluid with the force due to the
global potential acting on the chain $U = U_{\textsf{bend}} +
U_{\textsf{conn}} + U_{\textsf{solv}}$, and the Brownian force per
unit length $\etav$, with correlations $\la \etav(s,t)
\etav(s',t') \ra = 2 \kB T \zetav \delta(t-t') \delta(s-s')$.
$U_{\textsf{bend}} = kT L_p \int_0^L ds
\partial \vec{u}(s)/ \partial s $ is the bending potential related to the chain
stiffness, $U_{\textsf{conn}}$ is the connector potential,
$U_{\textsf{solv}}$ describes the monomer-monomer interaction, and
the Brownian term accounts for the rate of fluctuating momentum
transfer due to the random collisions of small solvent particles
with the chain.

The potential $U_{\textsf{solv}}$ chosen to account for the effect
of poor solvent is an exponential attractive potential with an
excluded volume repulsive term: \beqa \nonumber U_{\textsf{solv}}
= - \int ds \int ds' \tilde{u}_0 \exp \left(
-\frac{|\rv(s)-\rv(s')|}{R_{\textsf{attr}}} \right)
\\ + \int ds \int ds' \tilde{u}_0
\left( \frac{\sigma}{|\rv(s)-\rv(s')|} \right)^{12}, \eeqa where
$\tilde{u}_0$ defines the depth of the potential well,
$R_{\textsf{attr}}$ is the range of the attractive forces and
$\sigma$ is the radius of the excluded volume region. With this
potential, $u_0 = R_{\textsf{attr}} \tilde{u}_0$. Different
choices for the shape of the potential should not change the
stable and metastable configurations for the wormlike chain in
poor solvent, while they may modify the speed and the dynamics of
the collapse. Our choice of the exponential potential is related
to the physics of the condensation phenomenon. DNA and
polyelectrolytes collapse because of charge inversion and
counterion induced attraction \cite{Grosberg2002}. The attractive
forces are therefore due to highly screened electrostatic charges,
which typically show an exponential decay, mimicked by the
potential selected in this work.

The Brownian Dynamics code used in this work solves in
dimensionless, discretized form the Langevin equation of motion of
a linear wormlike chain of $N$ beads with positions
$\vec{R}_{1},\ldots,\vec{R}_{N}$, connected by $N-1$ connectors of
equilibrium length $a \equiv L/(N-1)$ with unit tangent vectors
$\vec{u}_{i} \equiv ( \vec{R}_{i+1} - \vec{R}_{i} )/a$. For most
of the simulations reported in this work, we use quadratic springs
as connectors, with a force constant equal to 100 $kT$, which
ensures a variation of the connectors' length within 30 $\%$. This
choice allows for longer time steps without interfering with the
spring relaxation time. We use a midstep algorithm
\cite{Grassia1996} to compute the bead trajectories generated by
the equation of motion
\begin{equation}
  \zeta_b \left \{ \frac{d\Rv_{i}}{dt} \! - \! \dot{\gamv}\cdot\Rv_{i} \right \}
  \! = \! \Fv_i
  \! = \! \Fv_{i}^{\textsf{bend}}
  \! + \! \Fv_{i}^{\textsf{elast}}
  \! + \! \Fv_{i}^{\textsf{solv}}
  \! + \! \Fv_{i}^{\textsf{rand}}
  \; . \label{Langevin-discrete}
\end{equation}
Here, $\zeta_{b} = \zeta a$ is a bead friction coefficient, the
bending force is: \beq  \Fv_{i}^{\bend} = \kappa/a
\sum_{j=2}^{N-1} \partial( \vec{u}_{j}\cdot\vec{u}_{j-1})/\partial
\Rv_i, \eeq the elastic force due to the spring is \beq
\Fv_{i}^{\textsf{elast}} = H \left[ ( |\uv_{i-1}| - a)\uv_{i-1} -
(|\uv_i| -a)\uv_{i} \right], \eeq the potential force related to
the effective interactions between the beads of the chain ---
repulsive within the hard core, attractive at short distance is
\beq \Fv_{i}^{\textsf{solv}} = \tilde{u}_0 \sum_{j \neq i} \left[
\frac {\exp(- \frac {|\bf{\Delta}\Rv_{ij}|} {R_{\textsf{attr}}}
)}{R_{\textsf{attr}} } - \frac{12 \sigma^{12}}
{|\bf{\Delta}\Rv_{ij}|^{13}} \right]
\frac{\bf{\Delta}\Rv_{ij}}{|\bf{\Delta}\Rv_{ij}|}\eeq where
$\bf{\Delta}\Rv_{ij} = \Rv_i - \Rv_j$,
%\beqa \nonumber \Fv_{i}^{\textsf{solv}} = \\ \tilde{u}_0
%\sum_{j=1, j \neq i}^{N} \left[ \frac {\exp(- \frac {|\Rv_i -
%\Rv_j|} {R_{\textsf{attr}}} )}{R_{\textsf{attr}} } - \frac{12
%\sigma^{12}} {|\Rv_i - \Rv_j|^{13}} \right] \frac{\Rv_i -
%\Rv_j}{|\Rv_i - \Rv_j|}, \eeqa
and $\Fv_{i}^{\textsf{rand}}$ is a
random force.

Another relevant issue in the simulations is the accurate
description of the high curvature of the collapsed states. The
discretized computational model for the wormlike chain retains its
validity when subsequent connectors are very nearly parallel:
%which requires that the angle between them is relatively close to
%180 $\angdegree$:
to satisfy this condition, a large number of connectors within the
characteristic length scale of the collapsed structure is needed.
The typical size of the torus and of the racquet heads are of the
same order of magnitude of $L_p$. A good accuracy in the
description of their curvature requires a large number of
connectors per persistence length; for all the simulations
presented here, we use 20 connectors per $L_p$.

We also performed simulations with a bead-rod model
\cite{Pasquali2001a,Pasquali2002c}, which ensures local
inextensibility of the worm-like chain. The computational cost for
this model is however significantly higher ($\sim$ 10-50 times)
than that for a bead-spring chain model, mainly because a smaller
time step is required for convergence. While performing most of
the simulations using the bead-spring model, we compared the
results for a small number of conditions with the bead-rod model,
as discussed in Section \ref{res4}. In that Section we also
compare some of the results with that obtained using a bead-spring
phantom chain, i.e., without excluded volume interactions.

For simplicity, we use isotropic drag in the simulations, and we
neglect hydrodynamic interactions between segments of the chain.
Including such interactions would likely give minor corrections to
the evaluation of the time scales of the phenomena, but ---
because at the beginning of the simulations the chains are locally
straight, and because after the collapse the hydrodynamic
interactions would be dominated by the attractive forces --- it
would not change the kinetic pathways and the relative stability
of the intermediate states.

\section{Results}

This section is organized as follows: Section \ref{res1} reports
qualitatively the results of our simulations, describing the
various possible intermediate configurations and the typical
pathways. Section \ref{res2} presents a more quantitative
description of the collapsed states in term of the correlation
function between tangent vectors along the chain and of the energy
levels associated with each different collapsed state, which
confirm that the toroid is the stable configuration, and show how
the collapse proceeds through a cascade of subsequently more
energetic favorable metastable states --- the various multiple
headed racquets shapes, as predicted from theory
\cite{Schnurr2002}. The decay rate of uncollapsed states is
investigated as a function of a proper combination of $L$, $L_0$
and $L_p$ in Section \ref{res3}, together with a statistical
analysis of the preferred collapsed configurations. Section
\ref{res4} provides a comparison between the results obtained with
the different computational models (phantom chain and bead-rod
chain). Some preliminary results of the effect of shear flow on
the kinetics and pathways of collapse are presented in Section
\ref{res5}.
\begin{figure}
\includegraphics[height=8.2cm,width=8.2cm]{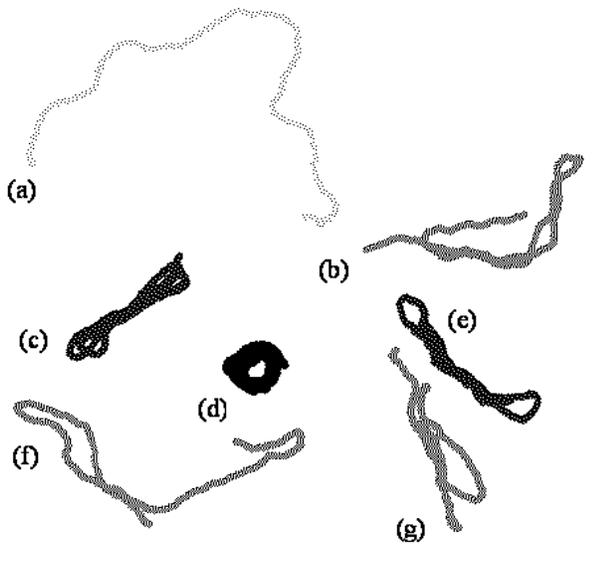}
\caption{Typical configurations of an ensemble of collapsing
chains at an intermediate time ($L/L_p = 8$, $u_0 = 1.25 kT/a$).}
\label{global view}
\end{figure}

%Throughout this paper, we will use $kT$ to normalize
%the energy, the persistence length $L_p$ as the natural length
%scale and the rotational diffusion time of a stiff chain of length
%$L_p$, i.e., $\tau_{\textsf{diff},L_p} = \left(\zeta
%L^3_p\right)/(72 kT)$ as characteristic time scale.

\subsection{Molecule configurations and collapse pathways}
\label{res1}

Figure \ref{global view} shows a typical snapshot of an ensemble
of semiflexible chains during collapse. The 3-D structures of the
molecules are shown here as 2-D projections on a convenient plane.
At a given time molecules coexist in different configurations,
from completely uncollapsed (molecule (a)), to partially collapsed
(molecules (b), (f) and (g)), to higher order racquet (molecules
(c) and (e)), to toroid (molecule (d)).

\begin{figure}
\includegraphics[height=2.35in,width=3.46in]{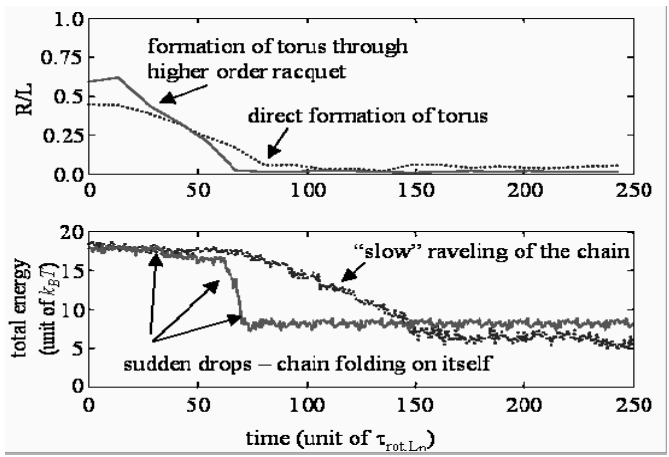}
\caption{Comparison between time series of end to end distance $R$
and total energy for direct formation of toroid and formation
through intermediate states.} \label{path_to_toroid}
\end{figure}

The fraction of collapsed molecules increases monotonically with
time, confirming that collapsed states are energetically favorable
in a bad solvent. Many other ensembles with different values of
$L/L_p$ and $L_0/L_p$ show a similar behavior. However, if the
attractive forces are not strong enough or the bending stiffness
is too high there is no evidence of collapse and the open
conformations persist in time. As already observed in
\cite{Schnurr2000}, a sequence of collapse can evolve following
two different generic patterns. The first pattern involves the
direct formation of a torus from the open conformation, when the
two ends of the chain meet while the tangent vectors are locally
parallel. In this instance, the chain collapses without formation
of any intermediate, the end to end distance drops quickly to a
value of the order of the torus dimension, and then fluctuates as
the chain keeps folding into the final structure. Meanwhile the
total energy of the chain keeps decreasing monotonically till
reaching the equilibrium value. Conversely, the second pattern
involves the formation of intermediate metastable states:
initially a single racquet-headed shape which folds quickly into a
short-lived, 3-heads racquet configuration, and then rearranges as
a toroid via a subsequent folding. Figure \ref{path_to_toroid}
shows the two different pathways. We notice here that the direct
formation of a torus is a more unlikely pattern in 3-D --- and in
fact observed with scarce frequency --- than in 2-D. Whereas in
2-D the relative orientation of the two ends is described by a
single angle, two angles are required in 3-D and both need to be
close to $2\pi$ for the direct formation of a torus to take place.
Most of the chain will therefore initially form a single headed
racquet, as described below \cite{movies}.

\begin{figure*}
\includegraphics[height=3.36in,width=6.01in]{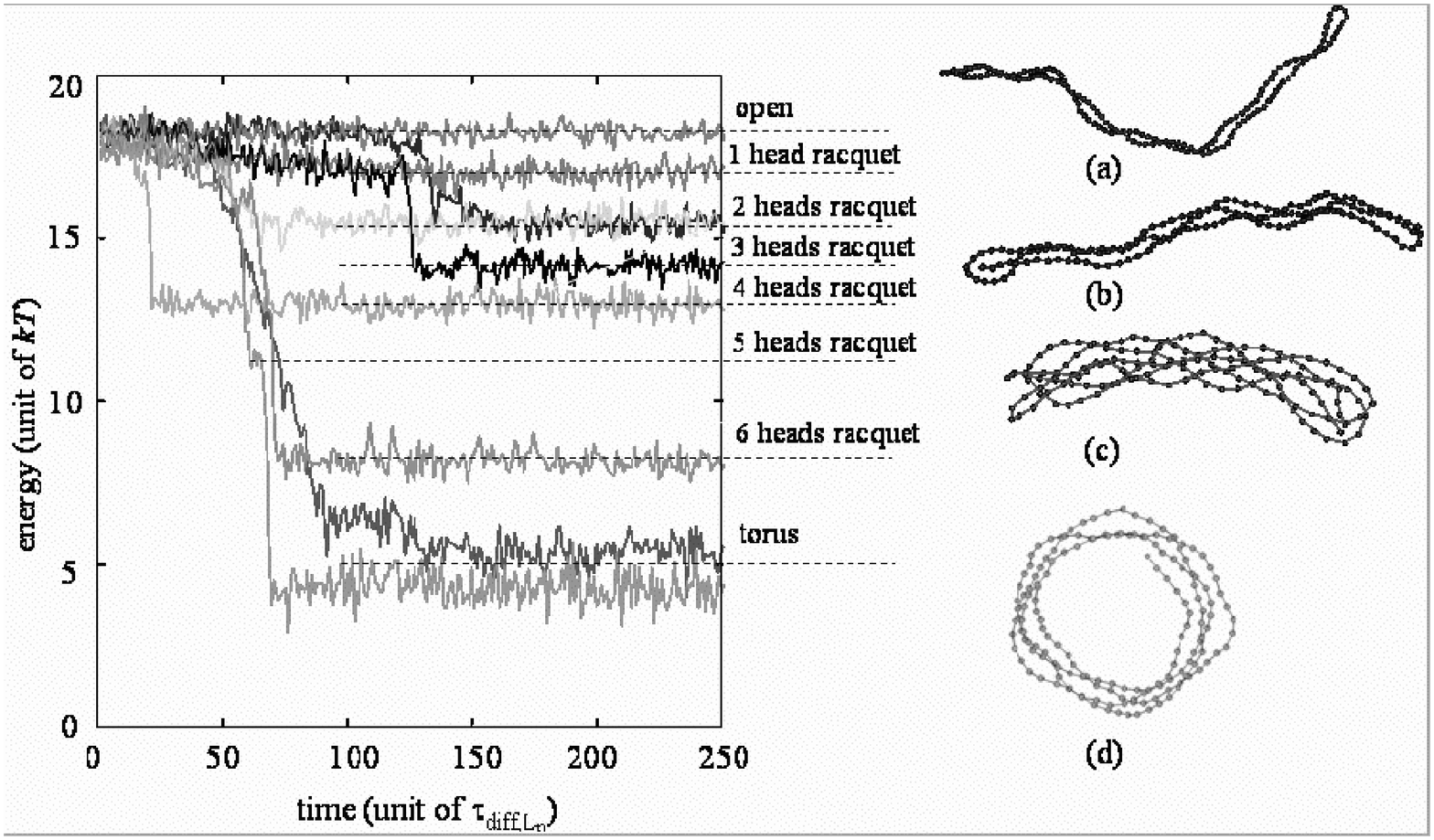}
\caption{Time series of total energy for an ensemble of molecules,
with labels of the corresponding configurations and actual shape
of (a) 1 head (b) 2 heads (c) 5 heads racquets and (d) torus.}
\label{time_series}
\end{figure*}

The time evolution of the total energy of an ensemble of
collapsing chains is shown in Figure \ref{time_series}. The total
energy $U$ of the chain during collapse is defined as the sum of
the bending energy $U_{\textsf{bend}}$  and the interaction energy
along the chain $U_{\textsf{solv}}$. In the discretized form,\beq
U = kT L_p  \sum_{i=1}^{N-2} \cos^2 \theta_i + \sum_{j=1}^{N}
\sum_{k=j+1}^N u_{\textsf{solv},jk}, \eeq where $\theta_i$ is the
bond angle and $u_{\textsf{solv},jk}$ is the interaction energy
evaluated between the beads $j$ and $k$ along the chain.

The time series in Figure \ref{time_series} reports most of the
possible evolutions observed in our simulations. Few molecules
remain in the uncollapsed states for the whole duration of the
run, equivalent to $250 ~ \tauLp$. For these molecules, the
thermal fluctuations never lead two branches of the chain close
enough together to initiate the collapse. The other molecules
typically reach lower energy states by collapsing in sequentially
higher order metastable conformations. The energy levels for each
racquet-head shape of increasing order are well separated, and
there is a rather large gap from the highest order racquet
observed ($6^{\rm{th}}$ order) and the stable tori, as expected
from theory \cite{Schnurr2002}. Most of the chains initially form
a single head racquet: after the first monomer-monomer contact and
the formation of the racquet head,  the two sides of the chain
tend to become parallel and form a neck region, which ensure the
maximum number of monomer-monomer contacts without increasing the
bending penalty. The molecules then fold again on themselves,
forming higher order shapes. Depending on the way in which the
molecule folds back on itself, the next configuration can be a
2-heads or a 3-heads racquet shape. Figure \ref{time_series} shows
examples of both pathways. Successive folding leads to higher
order racquets. The transition between each single metastable
conformation is very rapid and driven by the deterministic
attractive and bending forces. In contrast, the initiating event
for each successive collapse is related to the Brownian motion of
the molecule, and therefore the time interval between folding
events is extremely variable, and completely random for the
simulations that we have performed. In the simulations of longer
chains ( $L = 10 ~ L_p$ ) we observe combinations of the
conformational elements described above, e.g., chains partially
folded in a racquet shape at one end while still uncollapsed at
the other. Those conformations have intermediate energies between
the well-defined levels corresponding to racquets and tori and
quickly disappear in favor of more compact and ordered structures.
In contrast, the metastable high order racquets are extremely
long-lived, and the energetic barrier for reaching the equilibrium
shape is of the order of several $k_B T$. While the partial
unfolding of a torus in a racquet is never observed, simulations
show the formation of a torus via metastable configurations, and
this, together with the lower total energy shown by the torus (see
Figure \ref{time_series}), confirms once more that the latter is
the equilibrium configuration.

\subsection{Quantitative description of collapsed states} \label{res2}

The configurations and the shapes of semiflexible molecules in bad
solvent can be studied through direct visualization of the
collapsing chain. While this method is a valuable way to gather
information on the kinetics and pathways of collapse, it is time
consuming and not quantitative. We therefore define a spatial
correlation matrix $\bf{M}$ to analyze in a more systematic way
the collapsed shapes. The elements of $\bf{M}$ are defined as
$M_{ij} = \vec{u}_i \cdot \vec{u}_j$. With this definition,
$\bf{M}$ is symmetric; a perfectly straight rod has $M_{ij} \equiv
1 ~ \forall i,j$, as all the connecting normalized vectors have
the same direction and orientation. An uncollapsed semiflexible
chain with persistence length $L_p$ will have $M_{ii} \equiv 1$,
while the off-diagonal terms $M_{ij}$ would be smaller and
decaying exponentially with the distance along the chain between
the connectors $i$ and $j$.

When the chain collapses, the actual shape assumed by the molecule
can be inferred from the spatial correlation between connectors.
The spatial correlation matrix for a single head racquet, for
example, will be a block matrix: one block will show values close
to $1$, corresponding to the almost straight filament in the neck
of the racquet, followed by a region with rapidly decreasing
values from $1$ to $-1$ --- the racquet head --- and by another
block with values close to $-1$, indicating the other filament of
the neck , with same direction but opposite orientation.

The spatial correlation matrix can be simplified into a spatial
covariogram for the molecule. For standard statistical analysis,
the covariogram is defined as \beq C(h) =  1/N(h) \sum_{j=1}^N
\sum_{i=1}^N \{Z_i - \mu\} \{Z_j - \mu\} \eeq where $h$ is the
distance (in time or in space) between the observations $Z_i$ and
$Z_j$, $N(h)$ is the number of observations at a distance $h$ and
$\mu$ is the expected average value of the observation. In the
present analysis, for each molecule we construct the spatial
covariogram $C_{\textsf{sp}}$ assuming $Z_i = \vec{u}_i$ and $\mu
= \vec{0}$. We therefore obtain a scalar function of the distance
along the chain $s$, bounded between $+1$ and $-1$, which contains
all the relevant information about the shape of the molecule.

\begin{figure*}
\includegraphics[height=10.8cm,width=16.9cm]{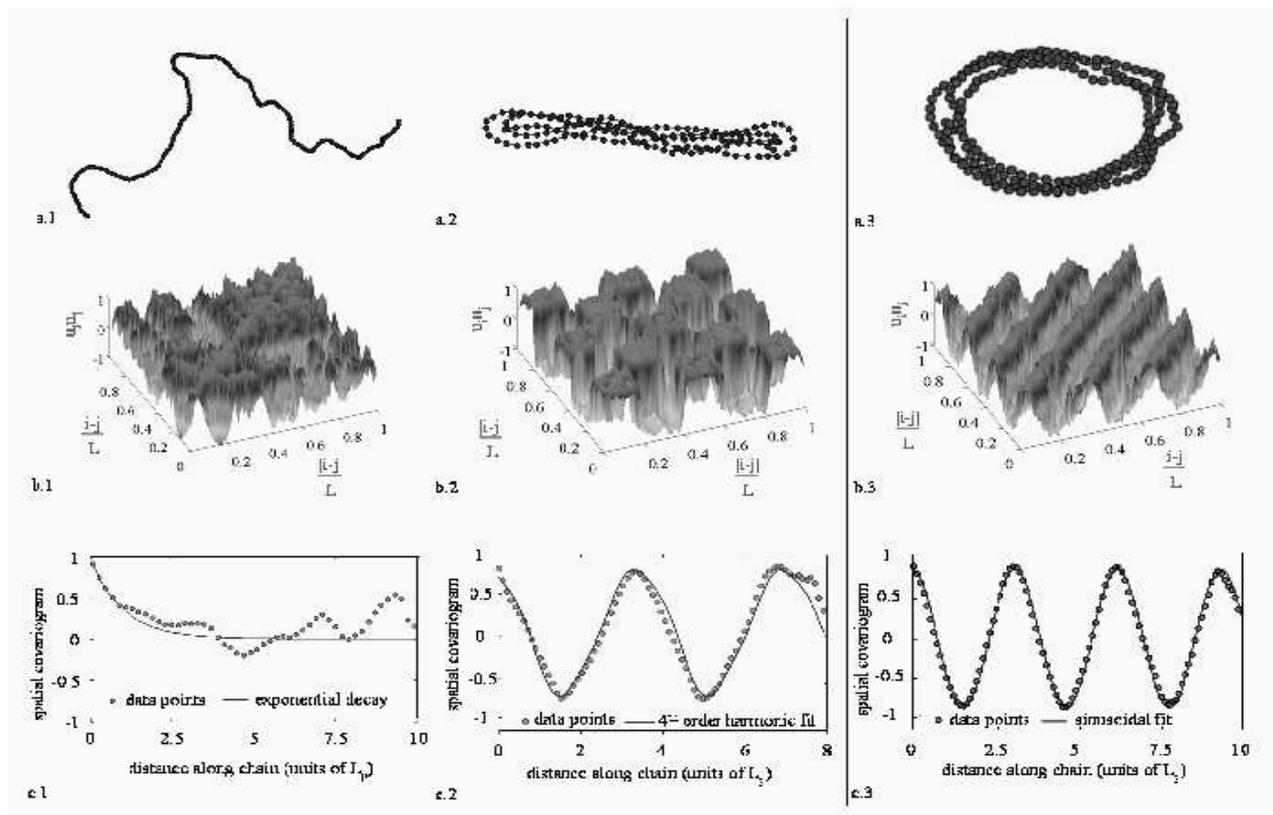}
\caption{(a) Actual shape, (b) correlation matrix and (c) spatial
covariogram (with fitting curve) for (1) an uncollapsed molecule,
(2) a metastable racquet-head shape and (3) a stable torus.}
\label{shape_matrix_covar}
\end{figure*}

Figure \ref{shape_matrix_covar} shows three relevant examples of
this type of analysis: the actual shape of the semiflexible
molecules is shown in comparison with the spatial correlation
matrix and the spatial covariogram. For a uncollapsed chain
(Figure \ref{shape_matrix_covar}-a), the matrix shows randomly
alternating regions of positive and negative correlation along the
length of the chain, corresponding to the typical open
configuration of a semiflexible chain in good solvent. The
corresponding covariogram shows how the connectors correlation
decays from 1 to values around 0 along the length of the chain;
the decay is exponential at short distances and related to the
stiffness of the chain, i.e., $C_{\textsf{sp}} = \exp (- s/L_p) $.
At longer distances, the correlation randomly oscillates between
negative and positive values.

For a racquet shape with multiple heads, $\bf{M}$ has a structure
of alternating blocks with strong positive and strong negative
correlation, as shown in Figure \ref{shape_matrix_covar}-b. The 5
straight filaments in the neck of the racquet correspond to these
5 blocks in the matrix; the size of the blocks is approximately
the same and indicates the length of the racquet neck. The
covariogram shows 4 zero-crossing, each corresponding to a head of
the racquet, and can be fitted quite accurately with a
$4^{\textsf{th}}$ order harmonic; the parameters of this fit are
related to the curvature of the racquet heads.

For a torus, the spatial correlation matrix appears completely
different (Figure \ref{shape_matrix_covar}-c), showing a structure
with diagonal bands. In fact, the correlation between pairs of
connectors at the same distance is the same, regardless of their
position along the chain; this translates into bands along the
diagonals of the matrix. The distance between a diagonal with
values close to $1$ and one with values close to $-1$ corresponds
to the diameter of the collapsed torus. The same information can
be extracted from the covariogram, which is a perfect sinusoid:
the period of the sinusoidal function corresponds to $\pi$ times
the diameter of the torus, and the number of complete periods in
along the chain indicates the number of loops formed by the
collapsed molecule.

%The collapsed structure can be analyzed also by looking at the
%eigenvalues of the spatial correlation matrix. It can be shown
%that, regardless the size of the matrix, i.e., regardless the
%discretization level of the simulated chain,
%$\bf{M}_{\textsf{corr}}$ has 3 non-zero eigenvalues, which are
%always positive. Their normalized values represent the major axes
%of an ellipsoid describing the molecule configuration, but not the
%orientation. Therefore for a semiflexible chain in good solvent
%the 3 eigenvalues are

\subsection{Kinetics of collapse} \label{res3}

The fraction of collapsed semiflexible molecules in a bad solvent
increases monotonically with time, as already stated. We have
studied the effect of the relevant length scales of the
phenomenon, $L_p$, $L_0$ and $L$, on the kinetics of this
collapse. By systematically varying the total length of the chain
($L = 3 L_p$, $5 L_p$, $8 L_p$ and $10 L_p$), as well as the
strength ($\tilde{u}_0 = 0.25 kT/a^2$, $0.5 kT/a^2$ and $1
kT/a^2$) and the range ($R_{\textsf{attr}} = 5/40 L_p$ and $1/6
L_p$) of the attractive forces, we have explored the effect of
these parameters on the decay rate of uncollapsed molecules. We
define the time to collapse $t_{\textsf{coll}}$ as the time
required by the molecule to form the first metastable collapsed
structure. We obtain from simulations the average value of
$t_{\textsf{coll}}$ in an ensemble of collapsing chain under each
condition, normalized with $\tauLp$. The inverse of
$t_{\textsf{coll}}$ is taken as a measure of the decay rate
$\textsl{D}_{\textsf{coll}}$ of uncollapsed molecules. The
distribution of the times to collapse follows the expected
lognormal distribution, typical of such dynamical processes, with
a peak close to the average value and a long tail.

\begin{figure}
\includegraphics[height=2.56in,width=3.17in]{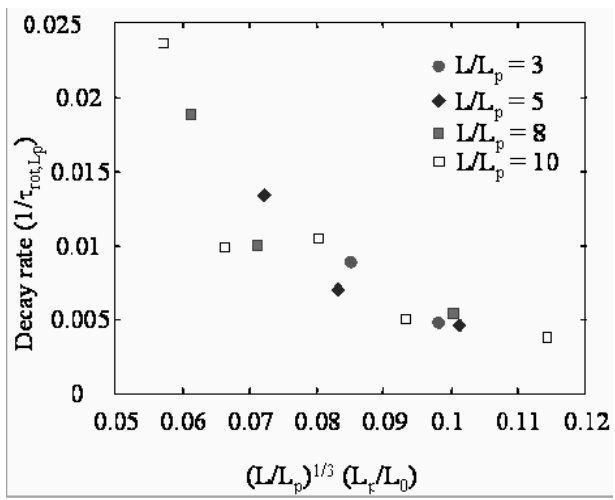}
\caption{Masterplot of the decay rate of uncollapsed molecules.}
\label{decay_rate}
\end{figure}

At a given value of $L/L_p$, the decay rate increases, i.e., the
molecules collapse faster, for decreasing values of $L_0$, i.e.,
for increasing values of $u_0$. Similarly, the decay rate
increases for longer chains at a given value of the attractive
potential and persistence length. We can collapse the values of
$\textsl{D}_{\textsf{coll}}$ on a single curve as a function of a
proper combination of the two dimensionless ratios, $L/L_p$ and
$L_0/L_p$, $\left(L/L_p\right)^{1/3} \left(L_p/L_0\right)$. Figure
\ref{decay_rate} shows the mastercurve for the decay rate.

Table \ref{table1} reports the statistics of chains conformations
for different values of $L/L_p$ at two different times, i.e., t1 $
= 100 \tauLp$ and t2 $ = 200 \tauLp$. At t1, the the single head
racquet is a conformation much more likely than the torus for all
the explored values of $L/L_p$.  This confirms that the direct
formation of a torus is an unlikely event, as discussed in Section
\ref{res1}, and most of the chains collapse through a sequence of
folding events. At t2, the percentage of molecules in the 1-head
racquet configuration does not vary substantially with respect to
t1, because --- while uncollapsed molecules fold to form new
1-head racquets --- some of these configurations further collapse
in higher order metastable racquets and stable tori. The number of
chains collapsed in tori or multiple head racquets monotonically
increases within the observed span of time. For shorter chains the
likelihood of tori and multiple racquets is similar, while the
metastable states seem to be initially more favorable for longer
chains ($L/L_p = 8$ and $L/Lp = 10$). Overall, the results show
that the intermediates conformations in the collapse phenomenon
are extremely long-lived: only $8.4 \%$ of the simulated chain has
reached the equilibrium, torus state after $200 \tauLp$.

\begin{table}
\begin{center}
\centering
%\begin{tabular}{|p{2.8cm}|@{  }p{1cm}|p{1cm}|p{1cm}|p{1cm}|p{1cm}|}
\begin{tabular}{|c||c|c||c|c||c|c||c|c|}
\hline %& & & & & \\
       & \multicolumn{2}{|c}{\textbf{$L/L_p = 3$}}
       & \multicolumn{2}{||c}{\textbf{$L/L_p = 5$}}
       & \multicolumn{2}{||c}{\textbf{$L/L_p = 8$}}
       & \multicolumn{2}{||c|}{\textbf{$L/L_p = 10$}}\\
       \cline{2-9}
\raisebox{1.5ex}[0cm][0cm]{\textbf{Conformation}}
  & t1 & t2& t1 & t2 & t1 & t2 & t1 & t2   \\
\hline %& & & & & \\
  {open chain}& $62.0 $ & $43.3 $ & $48.0 $ & $30.0 $ & $43.3 $ & $36.7  $ & $49.0 $ & $36.8 $\\
\hline %& & & & & \\
  {single-head }& $31.0 $ & $43.3$ & $36.0 $ & $42.5 $ & $33.4 $ & $30.0 $ & $24.5 $ & $23.7 $\\
\hline %& & & & & \\
  {multiple-heads } & $3.5 $ & $6.7 $ & $8.0 $ & $15.0 $ & $20.0 $ & $26.7 $ & $20.4 $ & $31.6 $\\
\hline %& & & & & \\
  {torus}& $3.5 $ & $7.9 $ & $8.0 $ & $12.5 $ & $3.4 $ & $6.7 $ & $6.1 $ & $7.9 $\\
\hline
\end{tabular}
\end{center}
\caption{Statistics of chain conformations for different $L/L_p$
at t1 $= 100 \tauLp$ and t2 $= 200 \tauLp$ ($\%$ values).}
\label{table1}
\end{table}

\subsection{Comparison of different computational models} \label{res4}

We have performed simulation of collapsing chains with different
computational models, in order to verify the results obtained and
the appropriateness of the model chosen. In particular, we have
performed simulations with a bead-rod model, where the chain
connectors are rigid rods rather than stiff elastic spring. As
noted in Section \ref{comp_mod}, the bead-rod model is a
constrained model which ensures local inextensibility of the
chain, but is computationally more expensive, due to more
stringent time step requirements for convergence. While local
inextensibility is fundamental when analyzing the polymer
contribution to the stress tensor \cite{Pasquali2001a}, its effect
on the pathways and dynamics of collapse is less relevant, as
confirmed by our results. In fact, we observe qualitatively
similar pathways to collapse in the simulations performed with the
bead-rod model. The preferred pathway includes the formation of
metastable racquet shapes, similarly to what observed in the
bead-spring simulations. The kinetics of the collapse are slower
for the locally inextensible model; we believe that this is due to
how the constraints are imposed, i.e., by projecting the random
forces onto the constraints, and therefore not allowing any
fluctuations along the chain. A more detailed comparison between
the collapse kinetics two model is prevented by the high
computational cost of simulating the constrained system.

\begin{figure}
\includegraphics[height=10.24cm,width=7.8cm]{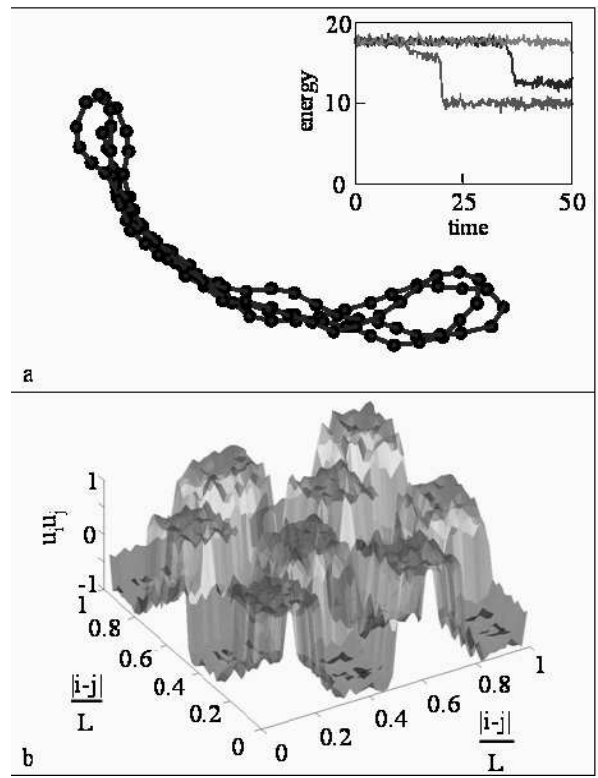}
\caption{(a) Metastable racquet-headed state for a phantom chain
and (b) corresponding correlation matrix. Inset in (a): energy
traces in time for collapsing phantom chains.} \label{phantom}
\end{figure}

We also perform simulations with a phantom bead-spring chain,
i.e., a chain with no repulsive hard core. Once more, we find that
the dynamics of collapse are qualitatively the same, although the
actual shapes of the torii and of the racquets formed by the
phantom chain are slightly different. The absence of hard core
repulsion allows in fact for tighter structures, as shown in
Figure \ref{phantom}. In particular, in the metastable multiple
racquets the neck for phantom chains appears much narrower than
for the chain where excluded volume interactions are taken into
account, while it is noticeable that the shape and the curvature
of the racquet head appears unchanged. The traces in time of the
energy level for the phantom chain show the features already
discussed in Section \ref{res1}: well defined energy states
corresponding to metastable structures, with rapid transitions
between them (see Figure \ref{phantom}, inset). However, the
average life of metastable states appear slightly shorter than in
the complete model: intuitively, this is due to the wider range of
motion of the phantom chain, which permits folding paths otherwise
prevented by the excluded volume interactions.

\subsection{Effect of shear flow} \label{res5}

We discuss here the results of some preliminary work on the
effects of steady shear flow on the collapse dynamics of
semiflexible chains, In particular we monitor the decay rate
$\textsl{D}_{\textsf{coll}}$ at different flow strengths, and we
compare the pathways of collapse with those at equilibrium.
Flexible chains expand in shear flow, i.e., the average end to end
distance increases while the molecules tend to orient with the
flow; the effect of the flow is therefore counteracting that of
the attractive forces, which tend to form compact globula. Stiff
semiflexible chains, with $L_p \geq L$ shrink in shear flow, due
to a buckling instability \cite{Montesi2003a}; such rigid chain
however would not collapse to form tori and multiple racquets in
bad solvent, as the open, rod-like shape is stable at equilibrium
\cite{Schnurr2002}. In this work, we consider the collapse
dynamics of less stiff semiflexible molecules: is not clear
\textit{a priori} what to expect for the collapse dynamics under
shear flow.
\begin{figure}
\includegraphics[height=2.56in,width=3.17in]{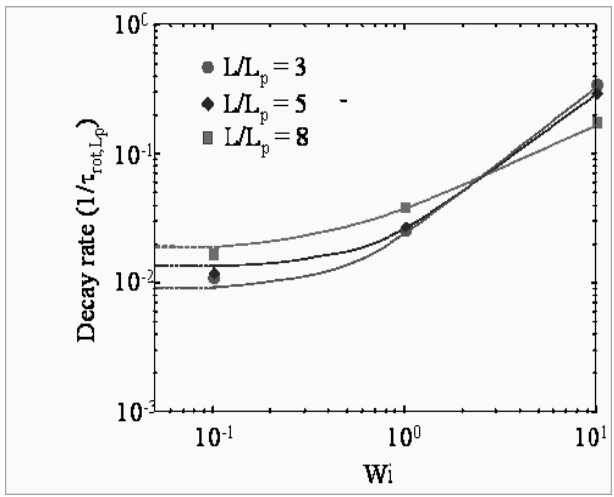}
\caption{Plot of the decay rate $\textsl{D}_{\textsf{coll}}$ vs.
Wi for $L/L_p =$ 3, 5, 8, $\tilde{u}_0 = 1 kT$ and
$R_{\textsf{attr}}$  = 10/3. } \label{decay_shear}
\end{figure}
\begin{figure}
\includegraphics[height=2.04in,width=2.84in]{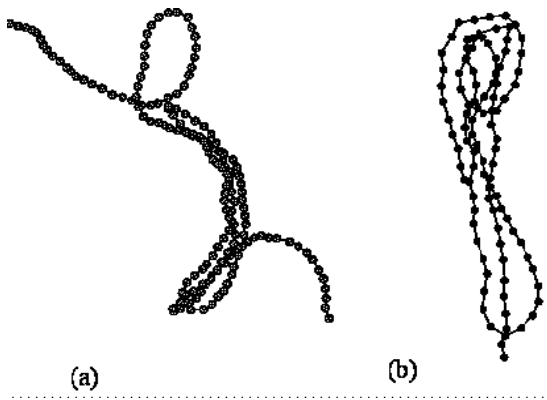}
\caption{Collapsed configurations of semiflexible chains in shear
flow at Wi $= 10$; (a) $L/L_p = 8$, (b) $L/L_p = 5$.}
\label{shapes_shear}
\end{figure}
We show here that strong enough shear flow speeds up sensibly the
collapse kinetics, increasing the likelihood for the two ends of
the chain to meet. Figure \ref{decay_shear} show the trend of the
decay rate as a function of Wi, where Wi $= \dot{\gamma}
\tau_{\textsf{diff},L_p}$, for 3 different values of $L/L_p$. At
Wi $= 0.1$, the decay rate does not change much with respect to
the equilibrium value, and there is not a clear trend among the
different $L/L_p$. At Wi $= 1$, the increase in decay rate becomes
instead significant, as the average time to first contact halves;
for Wi $= 10$, the effect of shear flow is completely dominating
and the decay rate under the 3 different conditions is almost
equivalent. Under shear flow, the semiflexible molecules undergo a
sequence of compressions and extensions while tumbling in the
plane of shear; in the compression region, the molecule tends to
shrink, therefore raising the probability that two sections far
apart along the chain backbone come in contact. The attractive
forces then trap the molecule in the partially collapsed
configuration, preventing it from straightening back in the
extension region. However, this continuous tumbling of the chain
and alternation of extension and compression does not allow the
formation of the higher order, compact structures observed at
equilibrium. In particular, the shear flow seems to inhibit the
formation of the torus, which was never observed in the high shear
flow simulations. Figure \ref{shapes_shear} shows two typical
configurations observed in strong shear flow; while they show the
basic features of the metastable racquets, the structures are less
compact and do not reach the same level of energy of the
equilibrium conditions.

\section{Discussion}

The dynamics of collapse of semiflexible molecules in bad solvent
has been investigated in the literature. Direct, visual
experimental evidence of the formation of torii and high order
racquets has been obtained studying biopolymers with sufficiently
large persistence lengths \cite{Baeza1987,Martin2000}. Theoretical
work and Monte Carlo simulations
\cite{Bloomfield1997,Noguchi1996,Schnurr2002} have confirmed that
the torus is the stable conformation, but intermediate, metastable
racquet shaped conformations exist and are long lived. Previous
Brownian Dynamics simulations in 2-D \cite{Schnurr2000} have
firstly suggested the possible general pathways to the dynamics of
collapse.

The present work provides new insights on the collapse of
semiflexible chains, confirming the suggested pathways through 3-D
simulations, and showing the effect of the two relevant
dimensionless ratios $L/L_p$ and $L_0/L_p$ on the kinetics of
collapse. The time series of the total energy of each chain show
clearly the fast transition between well defined energy levels,
corresponding to different metastable intermediate shapes, and the
final, stable torus. We also showed that two possible pathways
towards the stable conformation are possible: direct formation of
a torus and successive folding of the chain in progressively
higher order racquets. The latter one appears to be more probable
for the range of parameters investigated. The decay rate of
uncollapsed states for all the different conditions of the
simulation can be plotted as a function of a proper combination of
$L/L_p$ and $L_0/L_p$. Further work is needed to confirm the
validity of this empirical scaling and verify it in a wider range
of parameters.

We also performed simulations with different computational models,
therefore confirming that the observed metastable conformations
and pathways are general and do not depend on the choice of the
model. However, the kinetics, i.e., the actual value of the decay
rate can be different: in particular, for a constrained bead-rod
model the kinetics of collapse appear slower, possibly due to the
way the Brownian noise is projected onto the constraints.

We show preliminary results on the kinetics and pathways of
collapse of semiflexible chains in shear flow. Strong enough shear
flow increases the decay rate of uncollapsed states, but prevents
the formation of the compact, well defined configuration observed
at rest, and seems to inhibit the formation of the torus. While
these results have been obtained in unbounded shear flow, future
work will consider the interaction of the collapsing molecules
with a solid surface, which could be attractive or repulsive for
the chain. Also, future possible direction of research include the
effect of local defects along the chain, which make the molecule
locally more flexible or stiffer.

The authors wish to thank Bernhard Schnurr and F. Gittes for
useful discussions, and R. Adam Horch for preparing the
visualizations of the collapsing DNA molecules.  This work was
partially supported by the National Science Foundation under grant
PHY99-07949, through award CTS-CAREER-0134389 and through the
Nanoscale Science and Engineering Initiative award EEC-0118007.
Computational resources were provided by the Rice Parallel
Computational Engineering Cluster (NSF-MRI-0116289) and by the
SARA Computing and Networking Services in Amsterdam through Vrije
Universiteit.

\bibliography{amontesi,complete}

\begin{thebibliography}{25}
\bibitem[{fri()}]{friction_coeff}
\bibinfo{note}{From slender body hydrodynamics, $\zeta = 4 \pi \eta_s /
  \ln(L/r)$, where $\eta_s$ is the solvent viscosity and $L/r$ is the aspect
  ratio of the molecule.}
\end{thebibliography}

\end{document}